# Observation of cluster magnetic octupole domains in the antiferromagnetic Weyl semimetal Mn$_3$Sn nanowire

Hironari Isshiki[1,2]*, Nico Budai[1], Ayuko Kobayashi[1], Ryota Uesugi[1], Tomoya Higo[1,2,3], Satoru Nakatsuji[1,2,3,4] and YoshiChika Otani[1,2,4,5]

[1]Institute for Solid State Physics, The University of Tokyo; Kashiwa, Chiba, 277-8581, Japan.
[2]CREST, Japan Science and Technology Agency (JST); Saitama, 332-0012, Japan.
[3]Department of Physics, The University of Tokyo; Bunkyo-ku, Tokyo, 113-0033, Japan.
[4]Trans-scale Quantum Science Institute, The University of Tokyo; Bunkyo-ku, Tokyo, 113-0033, Japan.
[5]Center for Emergent Matter Science RIKEN; Wako, Saitama, 51-0198, Japan.
*Corresponding author. Email: h_isshiki@issp.u-tokyo.ac.jp

**Abstract:**

The antiferromagnetic Weyl semimetal Mn$_3$Sn has attracted wide attention due to their vast anomalous transverse transport properties despite barely net magnetization. So far, the magnetic properties of Mn$_3$Sn have been experimentally investigated on micrometer scale samples but not in nanometers. In this study, we measured the local anomalous Nernst effect of a (0001)-textured Mn$_3$Sn nanowire using a tip-contact-induced temperature gradient with an atomic force microscope. Our approach directly provides the distribution of the cluster magnetic octupole moments in Mn$_3$Sn with 80 nm spatial resolution, providing crucial information for integrating the Mn$_3$Sn nanostructure into a memory device.

The antiferromagnetic Weyl semimetals Mn$_3$X (X = Sn, Ge) with a non-collinear spin structure in a Kagome lattice have been drawing significant attention since they exhibit anomalous Hall and anomalous Nernst effects (ANE), which are generally absent in antiferromagnets [1–4]. The anomalous transverse transport properties of Mn$_3$X can be explained by the cluster magnetic octupole moment [5]. The cluster octupole moment behaves as a macroscopic order parameter, like a ferromagnetic moment, which characterizes the momentum-space Berry curvature [6–8]. These functional antiferromagnets hardly exhibit a magnetic shape anisotropy due to their negligibly small demagnetizing field [9]. This property will bring shape diversity into spintronic devices. For instance, the nanowires must be magnetized along the wire-width direction in thermoelectric generation and heat flux sensing using an anomalous Nernst thermopile structure with a vertical temperature gradient [9–12]. With ferromagnetic nanowires, due to the shape anisotropy, we have to apply a finite external magnetic field along the wire-width direction for efficient voltage generation by the ANE. In contrast, when utilizing antiferromagnetic materials like $Mn_3X$, the nearly negligible demagnetizing field enables this without an external magnetic field [9,13]. The efficiency and the stability of the spintronic devices based on such antiferromagnets rely on the controllability of the cluster magnetic octupole moments in the nanostructures. Therefore, observing the octupole moments in the nanostructure is crucial for the practical application of $Mn_3Sn$.

This paper presents the first demonstration of visualizing the spatial distribution of the cluster magnetic octupole moments within a (0001) oriented polycrystalline $Mn_3Sn$ nanowire. We employ a recently developed method based on conventional atomic force microscopy (AFM) [14,15]. This technique involves establishing a tip-to-sample contact where the sample is heated by a heater, inducing a localized temperature gradient [16,17], and measuring the thermoelectric voltages due to the ANE at the wire's ends [17], as illustrated in Figs. 1(a) and (b). Thus, unlike the stray field measurements with the nitrogen-vacancy center magnetometry [18,19], our signals directly reflect the orientation of the local octupole moments. Similar methods that induce temperature gradients by lasers have been applied to the films and the microwires of the antiferromagnetic Weyl semimetals [20,21], as well as magneto-optical Kerr effect measurements [22,23]. However, these measurements suffer from low spatial resolutions. Our technique has a higher spatial resolution of 80 nm and is, therefore, more suitable for investigating nanowires. Furthermore, by combining a numerical simulation, we quantitatively estimated the local thermoelectric coefficients of the order of the $Mn_3Sn$ grain. We believe the information obtained helps evaluate and/or optimize the $Mn_3Sn$ nanowires in antiferromagnetic spintronic and spin-caloritronic devices.

We employed a DC magnetron sputtering method for growing a polycrystalline $Mn_3Sn$ film on a Si(380 $\mu$m)/$SiO_2$(500 nm) substrate [9]. The 50-nm-thick $Mn_3Sn$ film is sputtered from a $Mn_{2.7}Sn$ alloy target using the sputtering power of 60 W and the Ar gas pressure of 0.8 Pa. After the deposition, the film is transferred to the annealing chamber and annealed at 723 K for 30 mins in vacuum (< $1\times10^{-6}$ Pa). The scanning electron microscope-energy dispersive x-ray spectroscopy reveals that the composition of the film is $Mn_{3.12(2)}Sn_{0.88(2)}$, which is in the range where the $D0_{19}$ $Mn_3Sn$ is reported to be stable [1,24]. The result of the $\theta$-$2\theta$ scan of the X-ray diffraction for the films is shown in

Fig. 1 (c). We confirmed the single peak of (002) for the $Mn_3Sn$ layer, which indicates that most of the Kagome planes are parallel to the substrate, i.e., the (0001)-textured $Mn_3Sn$ layer. We estimated the crystal grain size 100-250 nm [9,19] from the cross-sectional TEM measurement.

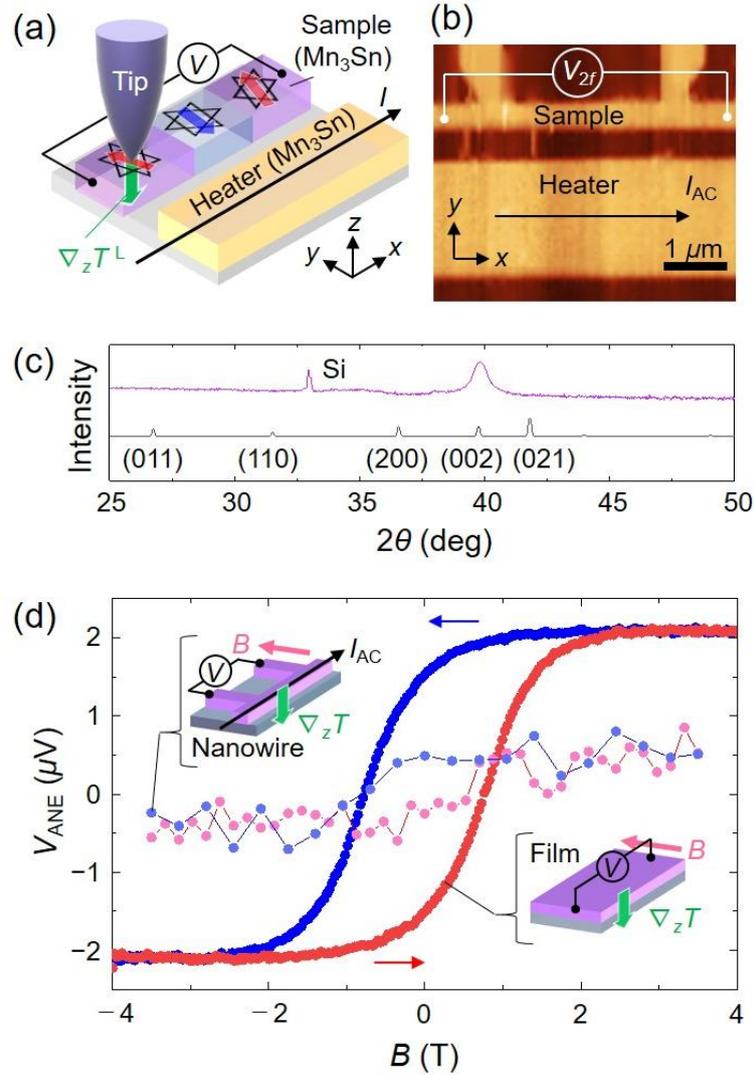

Figure 1: Concept of our experiment and the basic properties of the $Mn_3Sn$. (a) Conceptual illustration of the experiment. The sample temperature is controlled by Joule heating induced by a heating wire. Contacting a tip on the sample surface generates a local temperature gradient $\nabla_z T^L$ on the sample surface resulting in the voltage signal between the two ends of the wire, which is attributable to the local anomalous Nernst effect caused by the contact. Here, $\nabla_z T^L$ indicates the local temperature gradient. (b) Topography of the whole device consisting of the sample wire and heating wire with the illustration of the measurement configuration. (c) X-ray diffraction for $Mn_3Sn$ film. The experimental results for $Mn_3Sn$ and simulation on the material are shown by pink and black curves, respectively. (d)

Anomalous Nernst effects of the film (solid line) and the nanowire (solid circles) by sweeping the external magnetic field at room temperature. For the film, we create the temperature gradient by a heater. We apply the current to the nanowire sample to generate a temperature gradient.

We used the (0001)-textured $Mn_3Sn$ for the experiment below. Two parallel $Mn_3Sn$ wires were fabricated by electron beam lithography and Ar ion etching for the AFM experiments. One wire serves as the sample wire, 400 nm in width, and the other is a 1.9 $\mu$m wide heating wire. Their edge-to-edge separation is 600 nm. An AFM image of the device is shown in Fig. 1(b). We used an atomic force microscope CoreAFM from Nanosurf [25] to measure with a silicon cantilever Tap190Al-G (spring constant = 48 $Nm^{-1}$). Our device was placed on a homemade sample holder with a terminal for electrodes. We apply an alternating current (AC) to the heating wire with a frequency $f$. Joule heating increases the temperature of the sample wire by approximately 7 K, exhibiting oscillations at a frequency of $2f$. Here, we chose $f$ = 1043.43 Hz, which provides sufficient time for the heat to distribute thoroughly and still allows for a fast measurement. We scan the tip at the sample surface in contact mode with a loading force of 50 nN. The resulting tip-contact-induced thermoelectric voltages ($V_{2f}$) across the sample wire are detected using the standard lock-in technique. We acquired both AFM topography and a voltage $V_{2f}$ map simultaneously. The process required approximately 10 minutes to generate a set of images containing 400 pixels × 150 pixels. The details of this technique are described in our previous works [14,15]. All measurements were performed in the atmosphere at room temperature.

Initially, to confirm the behavior of the magnetic octupoles and their domains under in-plane magnetic field, we measured the ANE for the $Mn_3Sn$ thin film before processing (5.2 mm × 2 mm × 50 nm) and its fabricated nanowire (3 $\mu$m × 550 nm × 50 nm: one similar to that for the AFM measurement) over a varying external magnetic field at 300 K. The results are shown in Fig. 1(d) including the measurement configurations for both devices. For the film, the temperature gradient (the magnetic field) was applied in the direction perpendicular (parallel) to the film using the method reported in the previous works [9,26]. The temperature difference between the thermocouples placed on top and bottom of the sample, including the substrate, was obtained to be ~1.1 K, roughly suggesting the heat flux flowing perpendicular to the film of ~1 $W/cm^2$ [9,27]. Moreover, we adopted the method recently developed by Leiva et al. [28] for the nanowire. We applied an AC current $I_{AC}$ = 0.65 mA ($j_c$ = 2.4 × $10^{10}$ A/m) to the nanowire, which induces the out-of-plane temperature gradient in the whole wire. The anomalous Nernst voltage was lock-in detected as the second harmonic signal. According to the simulation using

COMSOL Multiphysics, the temperature of the nanowire is increased by ~ 23 K, which resulted in vertical induced temperature gradient $\nabla_z T$ ~ $2.7 \times 10^6$ K/m. One can see the saturation of the signals at ~ 2 T, with $V_{\text{ANE}} = 2.1$ $\mu$V for the film and $0.45 \pm 0.13$ $\mu$V for the nanowire. The hysteresis loop of the nanowire exhibits a well-defined remanent magnetization along the wire-width direction, indicating that the material retains a significant level of magnetization even in the absence of the external field because of the negligibly small shape anisotropy. In these measurements, the magnitude of the anomalous Nernst voltage $V_{\text{ANE}}$ is given by

$$V_{\text{ANE}} = l \cdot S_{\text{ANE}} \cdot \nabla_z T, \quad (1)$$

where $l$ is the length of the sample, $S_{\text{ANE}}$ is the coefficient for the ANE. The value of $S_{\text{ANE}}$ for the nanowire is estimated to be $0.056 \pm 0.16$ $\mu$V/K, nearly consistent with the one obtained in our previous works [9,27].

We show a magnified AFM image of the sample wire and the corresponding $V_{2f}$ map before applying an external magnetic field in Fig. 2(a) and (b), respectively. To obtain Fig. 2(b), we applied $I_{\text{AC}} = 3.4$ mA ($j_c = 3.6 \times 10^{10}$ A/m) to the heating wire. We set $V_{2f} = 0$ V at the substrate in Fig. 2(b) to remove the offset. The signal $V_{2f}$ at the sample wire is partially attributable to the local ANE in the textured $Mn_3Sn$ sample. The tip contact with the sample surface induces a local out-of-plane temperature gradient $\nabla_z T^{\text{L}}$. In our measurement configuration, the contributions of the in-plane temperature gradients are negligible since the magnitude of the x-component ($\nabla_x T^{\text{L}}$) is less than 2 % of $\nabla_z T^{\text{L}}$ according to the simulation and the y-component ($\nabla_y T^{\text{L}}$) does not produce anomalous Nernst voltage along the wire-length direction. The detectable anomalous Nernst voltage $V_{\text{ANE}}^{\text{L}}$ along the wire-length direction is given by,

$$V_{\text{ANE}}^{\text{L}} = \frac{1}{wt} \iiint S_{\text{ANE}} \cdot m_y \cdot \nabla_z T^{\text{L}} \, dxdydz \quad (2)$$

where $m_y$ is the unit vector of the cluster magnetic octupole y-component; $w$ and $t$ are the width and the thickness of the nanowire, respectively. The range of integral is the volume near the tip where the out-of-plane temperature gradient extends. The tip contact also induces a local temperature change $\Delta T^{\text{L}}$, that causes the Seebeck effect (SE) [15,29,30] at, for example, grain boundaries where the Seebeck coefficient discontinuously changes. Ignoring the spatial resolution, the Seebeck voltage $V_{\text{SE}}^{\text{L}}$ is given by

$$V_{\text{SE}}^{\text{L}} = \left(S_{xx}^{(A)} - S_{xx}^{(B)}\right) \cdot \Delta T^{\text{L}}, \quad (3)$$

where $S_{xx}^{(A)}$ and $S_{xx}^{(B)}$ are the Seebeck coefficients of adjacent two grains. Equation 3 holds by considering a thermopile structure created by two grains with slightly different Seebeck coefficients. Under the tip, a smaller temperature is induced compared to the equilibrium value, which generates a temperature difference between the tip and the end of the wire, creating a detectable voltage difference in our thermopile-like setup. Therefore, the signal shown in Fig. 2(b) results from the local ANE and SE. A more significant focus on spatial resolution will be set later in this work.

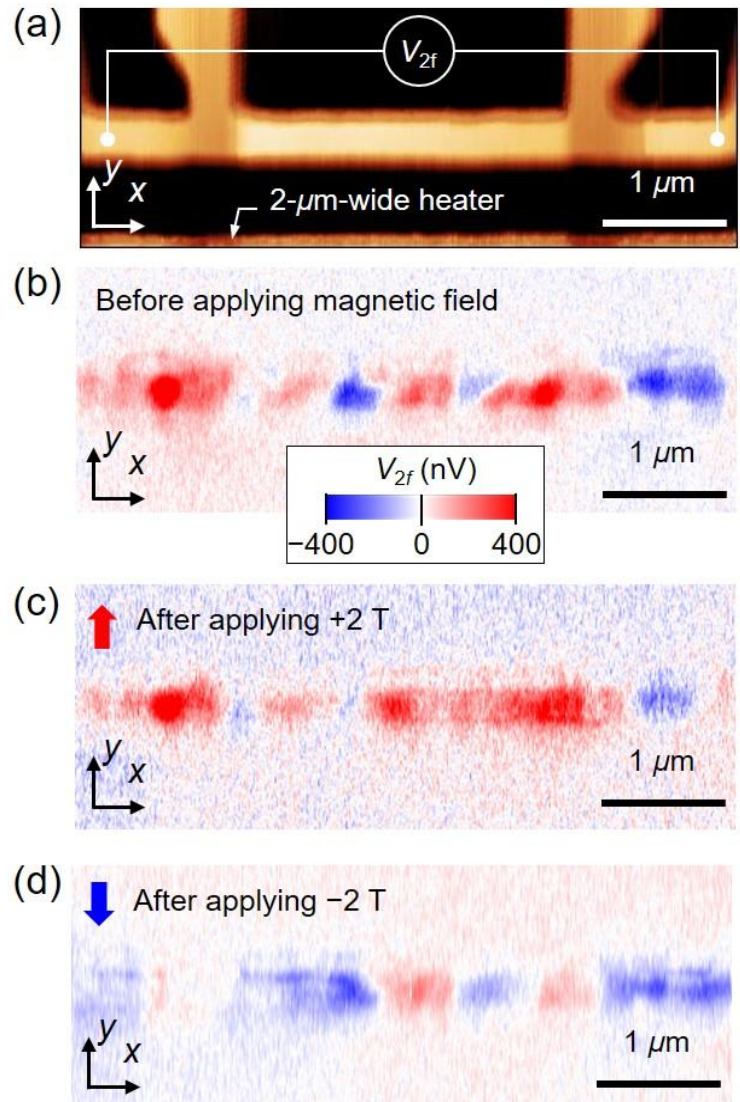

Figure 2: Thermoelectric voltage mapping on the $Mn_3Sn$ sample wire. (a) Magnified topography of the $Mn_3Sn$ sample wire. (b) The local thermoelectric voltage mapping on the $Mn_3Sn$ wire before applying an external magnetic field. (c), (d) the same but after applying positive and negative 2 T,

respectively. The scanning areas for (a)-(d) are identical. The voltage mappings are obtained simultaneously to the topography with the contact mode. The scan range is 5.4 $\mu$m × 2.0 $\mu$m (400 pixels × 150 pixels).

To observe the magnetic response of the Mn$_3$Sn, we employed a magnetizing process along the ±$y$-direction. The sample was positioned within an independent electromagnet, allowing us to control the magnetic field in the range of ±2 T. We smoothly swept the magnetic field strength from 0 T to 2 T in the positive or negative $y$-directions, subsequently returning it to 0 T. After the application of the magnetic field, we repositioned the sample back into the AFM at its original location. We then repeated the $V_{2f}$ mapping procedure at 0 T. The resulting $V_{2f}$ mappings after applying ±2 T are represented in Figs 2 (c) and (d), respectively. Notably, following the application of a positive (negative) magnetic field, we observed an expansion of the areas displaying positive (negative) $V_{2f}$ signals. This clearly illustrates the presence of a magnetic component within our signal.

We can numerically separate the magnetic (ANE) and non-magnetic (SE) signals from Figs. 2 (c) and (d) by having the following assumption: As magnetizing oppositely, the individual magnetic octupole moments reverse the direction while the positions of the magnetic domain boundaries remain unchanged. This assumption can be verified by the previous reports on polycrystalline Mn$_3$Sn films studied using the nitrogen-vacancy center with AFM [18,19]. Since the non-magnetic SE signal is unchanged between the positively and negatively magnetized sample, the average and the half difference between the $V_{2f}$ in Figs. 2 (c) and (d) give the SE and ANE signals, respectively. The averaging and subtraction were done after carefully aligning the shift of scanning positions. The results are shown in Figs. 3(a) and (b). We also subtract the non-magnetic signals in Fig. 3(a) from the $V_{2f}$ in Fig. 2(d), as shown in Fig. 3(c), which should indicate the ANE signal in the initial magnetic state. In Fig. 3(d), we show the line profile of the $V_{SE}$ and $V_{ANE}$ signals on the dashed lines in Figs. 3(a)-(c). We will discuss the details in Fig. 3 after showing the simulation of the temperature distribution change induced by the tip. We repeated the same measurements and analysis on another Mn$_3$Sn device. The results shown in the supplemental material are quantitatively very similar to Fig. 2(b)-(d) and Fig. 3(a)-(c).

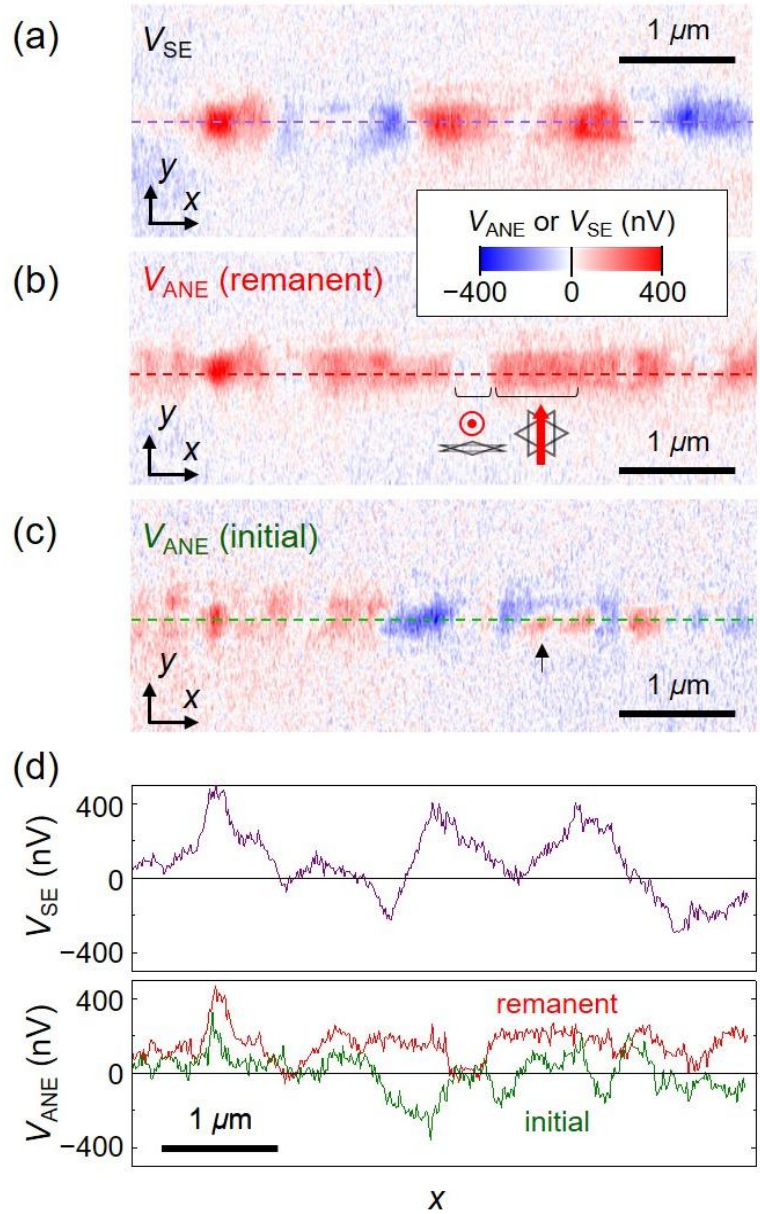

Figure 3: Numerically extracted non-magnetic and magnetic signals. (a) Non-magnetic signal. (b) Magnetic signal after applying 2 T. (c) Magnetic signal in the initial state. The averaging and the subtraction between the signals in Figs. 2(b)-(d) produce these images. The numerical processing was performed after carefully aligning the shift of scanning positions. (d) The line profiles on the dashed lines in (a)-(c). The green and red profiles represent the magnetic signals before and after applying 2 T (remanent).

For discussion on the quantitative thermoelectric coefficients and the spatial resolution of our technique, we need the temperature distribution induced by the tip contact in the sample wire. We performed numerical simulations on the geometry of our real device by using COMSOL Multiphysics [31]. According to the studies in scanning thermal microscopy [32–34], the dominant heat transfer mechanisms between the tip and the sample are air conduction and water meniscus [35]. In our model, these mechanisms are represented by a disc-shaped virtual material inserted in between the tip and sample with phenomenological parameters: the thermal conductance $G_c$ = 20 $\mu$W/K and the contact thermal radius $r_c$ = 30 nm (that should be comparable to the tip radius > 10 nm) [15]. The heat conductivities of $Mn_3Sn$, Si (tip and substrate), and $SiO_2$ are 12 [36,37], 130 [38], and 1.2 W/m/K [39], respectively. We set the initial temperature to 293.15 K and applied a current of 3.4 mA ($j_c$ = 3.6 × $10^{10}$ A/m) to the heating wire next to the sample. The simulated results are shown in Fig. 4(a) and (b), the temperature distribution in the *x-z* plane across the sample wire with and without tip contact, respectively. Before the tip contact, the temperature of the sample wire is ~ 299 K, which is almost homogeneous, as shown in Fig. 4(a). The tip contact drastically changes the landscape of the temperature distribution as shown in Fig. 4(b). The temperature change $\Delta T^L$ (~ 1.2 K on the average in the sample) extent of ~ 500 nm. The out-of-plane temperature gradient $\nabla_z T^L$ with tip contact in the same plane is shown in Fig. 4 (c). Remarkably, the out-of-plane temperature gradient is produced very locally: the extent in the *x*-direction is ~ 80 nm. Threfore, The integral in Eq. (2) should be taken in a cylindrical region with the radius of 80 nm in the $Mn_3Sn$ sample (the magnitude of $\nabla_z T^L$ is ~ 4×$10^6$ K/m on average in the cylindrical region). We found that the resolution of ANE (~ 80 nm) is much better than that of SE (~ 500 nm) in our technique, that is consistent to our previous experiments [14,15].

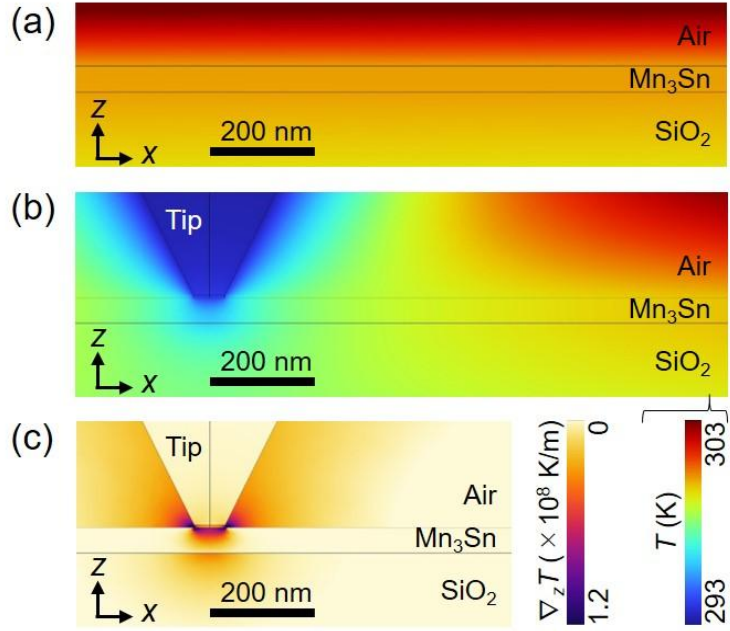

Figure 4: Simulated temperature distribution of the Mn$_3$Sn sample wire by COMSOL Multiphysics. (a) Temperature distribution in the *x-z* plane without tip contact. (b) Temperature distribution with tip contact. (c) Distribution of the out-of-plane temperature gradient with tip contact. The same geometry as the actual device has been set. The heating wire is behind the sample (Mn$_3$Sn) wire.

Figures 3(b) and (c) represent the distributions of the *y*-component of the cluster magnetic octupole domains in the Mn$_3$Sn nanowire with a spatial resolution of 80 nm. Before applying a magnetic field, as shown in Fig. 3(c), the $V_{ANE}$ becomes positive or negative depending on the position, which implies the randomly distributed octupole domains in the initial state. A small island of $V_{ANE}$ marked by black arrows in Fig. 3(c) is attributable to a grain of Mn$_3$Sn. After applying a magnetic field, as shown in Fig 3(b), the $V_{ANE}$ distribution changes, and the value becomes always positive. These results indicate that the octupole domains exhibit a remanent state along the wire-width +y direction. However, the $V_{ANE}$ signal is inhomogeneous, reflecting the presence of the grains with tilted Kagome planes about the *x*-axis, consistent with the broad peak of (002) in Fig. 1(c), *i.e.,* some grains with not-well-oriented Kagome planes. Using a typical value of $V_{ANE}$ ~ 180 nV in Figs. 3(b) and the simulated value of $\nabla_z T^L$ ~ 4×10$^6$ K/m, we obtained $S_{ANE}$ ~ 0.27 $\mu$V/K. This value is greater than the estimation by the conventional method shown in Fig. 1(d) but close to the reported one in the single crystal Mn$_3$Sn (~ 0.3 $\mu$V/K) [37,40]. The non-uniform ANE distribution per grain explains this tendency: some grains do not fully contribute to the ANE, probably due to their tilted Kagome planes, which decreases the thermoelectric efficiency from the intrinsic one. Using a typical value of $\left|V_{\text{SE}}^{\text{L}}\right|$ ~ 200 nV and Eq. (3), we obtain $\left(S_{xx}^{(A)} - S_{xx}^{(B)}\right)$ ~ 0.17 $\mu$V/K. As we have discussed before, due to the insufficient resolution, this value must be averaged across several grains, but the order of magnitude is consistent with the previous report, in which the $S_{xx}$ along $[2\bar{1}\bar{1}0]$ and $[01\bar{1}0]$ differ by 0.7 $\mu$V/K [36].

We observed the local anomalous Nernst effect in a (0001)-textured Mn$_3$Sn nanowire using the tip-contact-induced temperature gradient with 80 nm spatial resolution. Not like the magnetic imaging by the stray field measurement, our approach directly maps the distribution of the cluster magnetic octupole moments in Mn$_3$Sn. We visualized the octupole domains in the initial and remanent states in the nanowire, which is crucial information for the integration of Mn$_3$Sn. Our work provides a solid methodology to investigate the magnetic structures of the antiferromagnetic Weyl semimetal.

**Acknowledgement**

This work was partially supported by CREST (Grant No. JPMJCR18T3) from JST, JST-Mirai Program (Grant No. JPMJMI20A1), JSPS KAKENHI Grant-in-Aid for Scientific Research(C) (Grant Nos. 19H05629), Steel Foundation for Environmental Protection Technology.